\documentclass[prd,twocolumn,superscriptaddress,altaffilletter,showpacs]{revtex4}
\usepackage{amssymb,amsmath}

\def\be{\begin{equation}}
\def\ee{\end{equation}}
\def\bea{\begin{eqnarray}}
\def\eea{\end{eqnarray}}

\begin{document}
\title{Kerr black hole parameters in terms of red/blue shifts of photons emitted by geodesic particles}

\author{Alfredo Herrera--Aguilar} \email{aherrera@ifuap.buap.mx}
\affiliation{Instituto de F\'{\i}sica, Benem\'erita Universidad Aut\'onoma de Puebla,\\ 
Apartado Postal J-48, 72570, Puebla, Puebla, M\'exico.}
\affiliation{Instituto de F\'{\i}sica y Matem\'{a}ticas, Universidad Michoacana de San Nicol\'as de Hidalgo,\\
Edificio C--3, Ciudad Universitaria, CP 58040, Morelia, Michoac\'{a}n, M\'{e}xico.}

\author{Ulises Nucamendi} \email{ulises@ifm.umich.mx}
\affiliation{Instituto de F\'{\i}sica y Matem\'{a}ticas, Universidad Michoacana de San Nicol\'as de Hidalgo,\\
Edificio C--3, Ciudad Universitaria, CP 58040, Morelia, Michoac\'{a}n, M\'{e}xico.}
\date{\today}

\begin{abstract}
We are motivated by the recently reported dynamical evidence of stars with short orbital periods moving around the center of the Milky Way and the corresponding hypothesis about the existence of a supermassive black hole hosted at its center. In this paper we show how the mass and rotation parameters of a Kerr black hole (assuming that the putative supermassive black hole is of this type), as well as the distance that separates the black hole from the Earth, can be estimated in a relativistic way in terms of i) the red and blue shifts of photons that are emitted by geodesic massive particles (stars and galactic gas) and travel along null geodesics towards a distant observer, and ii) the radius of these star/gas orbits.

As a concrete example and as a first step towards a full relativistic analysis of the above mentioned star orbits around the center of our galaxy, we consider stable equatorial circular orbits of stars and express their corresponding red/blue shifts in terms of the metric parameters (mass and angular momentum per unit mass) and the orbital radii of both the emitter star (and/or galactic gas) and the distant observer.

In principle, these expressions allow one to statistically estimate the mass and rotation parameters of the Kerr black hole, and the radius of our orbit, through a Bayesian fitting, i.e., with the aid of observational data: the red/blue shifts measured at certain points of stars' orbits and their radii, with their respective errors, a task that we hope to perform in the near future. We also point to several astrophysical phenomena, like accretion discs of rotating black holes, binary systems and active galactic nuclei, among others, to which this formalism can be applied.

\end{abstract}


\maketitle

%%%%%%%%%%%%%%%%%%%%%%%%%%%%%%%%%%%%%%%%%%%%%%%%%%%%%%%%%%%%%%%%%%%%

\section{Introduction}

In the last decades astrophysical observations point out to dynamical evidence supporting the existence of a super massive black hole, called SgrA*, at the center of the Milky Way, as well as in the center of many other spiral galaxies \cite{Begelman,Shen,Ghez,Morris}. Nowadays, two teams of astronomers have managed to track the orbits of several stars orbiting around the center of our galaxy \cite{Eckart,Gillessen}.
These recent observational data allowed the aforementioned teams to estimate the putative black hole mass for SgrA* and the distance from Earth to its center, rendering the following values $M \sim 4.3 \times 10^{6} M_{\bigodot}$ and $R_0 \sim 8.3$ kpc. These researchers have computed these quantities by making use of a Newtonian approach, i.e., namely, they use a Keplerian central potential that assumes that the black hole mass is concentrated in a point-like object. Because of this reason, they just compute the mass of the black hole and the distance from Earth to the center of SgrA* and do not compute its angular momentum.

As a complementary work, the rotation parameter has been estimated and bounded using different indirect methods which lead to quite different but congruent results; for instance, when using high-frequency quasi periodic oscillations, computations render a value around  $a\sim 0.996 M$ \cite{aschenbach}, whereas when using flare emissions with a certain period, calculations lead to the following estimation $0.70 \pm 0.11 \, M \leq a \leq M$ \cite{trippe2007}.

With the aim of getting more precise values for the above referred parameters and quantities, and directly characterizing the assumed black hole hosted at the center of our galaxy, in the near future there will be new experiments to come: GRAVITY will track with more accuracy the orbits of stars around the centre of our galaxy \cite{gravity}, whereas the Event Horizon Telescope will focus on the black hole event horizon looking for traces of its shadow measuring signals in the infrared spectrum \cite{EHT}.

In 4D General Relativity, neutral rotating black holes are described by the Kerr solution and are completely characterized by just two physical quantities: the mass $M$ and the angular rotation parameter $a = J/M$, where $J$ is the black hole angular momentum. We shall use this metric in order to model the putative black hole hosted at the center of the Milky Way (SgrA*). Therefore, the aim of this paper is to provide a method for determining the $M$ and $a$ parameters of the Kerr black hole, as well as the distance from the black hole to the Earth, in terms of the directly measured red and blue shifts of photons emitted by massive particles (stars and gas) moving along geodesics around it, and the radius of their orbits. Thus, our approach is completely relativistic and allow us to compute the mass as well the rotation parameter of the Kerr black hole, in addition to the distance that separates the black hole from the Earth.

The paper is organized as follows: In Sec. II, we consider stationary axisymmetric metrics, their Killing vector fields and conserved quantities as well as the geodesic motion of massive particles on this family of space-times. In Sec. III, we introduce the Kerr metric, its Killing tensor field, and compute the corresponding Carter constant of motion. In Sec. IV, we consider photons which travel along null geodesics from the emisor (the massive particle that represents stars and orbiting gas/dust) to the detector (a far away located observer, the Earth for practical purposes). In Sec. V, we further compute the general expression for the red and blue shifts of these photons when taking into account the light bending due to the axisymmetric stationary gravitational field at the moment of detection by the observer. We further compute these shifts for the Kerr metric and particularize for circular and equatorial orbits of massive particles and photons that travel in the equatorial plane ($\theta=\pi/2$) in Sec. VI.

With these expressions at hand, we are in position of computing the mass, the rotation parameter of the source, i.e. a Kerr black hole, as well as the distance from it to the Earth in terms of the red/blue shifts that these photons experience and the radius of the emitter star orbits. We finally make some final remarks in Sec. VII, where we also discuss our results.

\section{Particles in Stationary Axisymmetric Spacetimes}

In order to achieve our aim, we shall first consider a massive test particle which follows a time-like geodesic path on a rotating
axially symmetric space-time. The most general metric for a space-time of this kind with two orthogonal planes reads:
\begin{equation}
\label{metric}
       ds^2 = g_{tt} dt^2 +  2 g_{t\varphi} dtd\varphi + g_{\varphi\varphi} d\varphi^2
       + g_{rr} dr^2 + g_{\theta\theta} d\theta^2,
\end{equation}
where we have chosen spherical coordinates $x^{\mu}(t,r,\theta,\varphi)$ as well as the gauge $g_{r\theta}=0$. Thus, all metric
components have the following dependence $g_{\mu\nu}(r, \theta)$ and $ \mu, \nu = t,r,\theta,\varphi$.

The metric (\ref{metric}) possesses two commuting Killing vector fields $[\xi,\psi] = 0$:
\begin{eqnarray}
\label{KillingT}
  \xi^\mu &=& (1,0,0,0)   \quad \textrm{time-like Killing vector field,}  \\
  \label{KillingR}
  \psi^\mu &=& (0,0,0,1)  \quad \textrm{rotational Killing vector field.}
\end{eqnarray}
Thus, the photons' {\it emitter} is a probe massive particle which geodesically moves around a rotating axisymmetric source in the
space-time represented by the metric (\ref{metric}) with a 4-velocity
\begin{equation}\label{velocity}
    U^{\mu}_{e} = (U^{t}, U^{r}, U^{\theta}, U^{\varphi})_{e},
\end{equation}
which is normalized to unity $u^\mu u_\mu=-1$, rendering the following relation
\begin{equation}
\label{mg}
-\!1\!=\!g_{tt} \left(U^t\right)^2\!+\!g_{rr}\left(U^r\right)^2\!+\!g_{\varphi\varphi}\left(U^\varphi\right)^2
\!+\!g_{\theta\theta} \left(U^{\theta}\right)^2\!+\!2g_{t\varphi}U^tU^\varphi.
\end{equation}

Due to the existence of the Killing vector fields (\ref{KillingT}) and (\ref{KillingR}) there are two
conserved quantities, the total energy and the angular momentum per unit mass at rest of the test
particle:
\begin{eqnarray}
  E &=& \frac{\bar{E}}{m} = - g_{\mu\nu} \xi^{\mu} U^{\nu} = - g_{tt}U^{t} - g_{t\varphi} U^{\varphi},
  \label{E}  \\
  L &=& \frac{\bar{L}}{m} = \quad g_{\mu\nu} \psi^{\mu} U^{\nu} = g_{\varphi t}U^{t} + g_{\varphi\varphi} U^{\varphi}.
  \label{L}
\end{eqnarray}

From these relations we obtain the expressions for $U^{t}$ and $U^{\varphi}$ in terms of the metric components
and the conserved quantities $E$ and $L$:
\begin{equation}
\label{Ut}
U^t =
\frac{Eg_{\varphi\varphi}+Lg_{t\varphi}}{g_{t\varphi}^2-g_{tt}g_{\varphi\varphi}},
\end{equation}
\begin{equation}
\label{Ufi}
U^{\varphi}=
-\frac{Eg_{t\varphi}+Lg_{tt}}{g_{t\varphi}^2-g_{tt}g_{\varphi\varphi}}.
\end{equation}
By substituting these 4-velocity components in the expression (\ref{mg}) we obtain
\begin{eqnarray}
&&g_{rr} (U^{r})^2\!+\!g_{\theta\theta} (U^{\theta})^2\!+\!1\!-\!\frac{E^2 g_{\varphi\varphi}\!+\!2E\,L\,g_{t\varphi}\!+\!L^2 g_{tt}}{(g_{t\varphi}^{2}-g_{tt}g_{\varphi\varphi})}\!=
\nonumber\\
&&g_{rr} (U^{r})^2\!+\!V_{eff}=0.
    \label{Ur2}
\end{eqnarray}
It is worth mentioning that this equation has the form of the energy conservation law for a non-relativistic particle with position
dependent mass moving in an effective potential $V_{eff}$ that depends on the conserved quantities $E$ and $L$ as well as on the metric
$g_{\mu\nu}$.
This effective potential must possess a maximum, implying that the following conditions must be obeyed for circular orbits (when $U^r=0$ in (\ref{Ur2})) \cite{Bardeen,Wilkins}:
\begin{equation}
\label{Vmin}
%, \qquad
V_{eff}=0,   \qquad \mbox{\rm and} \qquad V'_{eff}=0,
\end{equation}
where primes denote derivatives with respect to $r$.
The stability condition for these circular orbits leads to 
\begin{equation}
\label{Veff=0}
V''_{eff} < 0.
\end{equation}
Additionally, we can also restrict our analysis to equatorial orbits ($\theta=\pi/2$) implying that
$U^\theta=0$ and getting a further simplification.

\section{The Kerr black hole and its Killing tensor field}

The Kerr black hole family in Boyer-Lindquist coordinates is given by the metric (\ref{metric}) with the following components
\begin{eqnarray}
    g_{tt}\!=\!-\!\left(1\!-\!\frac{2Mr}{\Sigma}\right),\quad g_{t\varphi}\!=\!-\!\left(\frac{2Mar\sin^2 \theta}{\Sigma}\right),
    \quad g_{rr}\!=\!\frac{\Sigma}{\Delta},  \nonumber
\end{eqnarray}
\begin{equation}
    g_{\varphi\varphi}\!=\!\left(r^2\!+\!a^2\!+\!\frac{2Ma^2r\sin^2 \theta}{\Sigma}\right)\sin^2\theta,\qquad
    g_{\theta\theta}\!=\!\Sigma,
\end{equation}
where
\begin{eqnarray}
    \Delta = r^2 + a^2 - 2Mr\,,\qquad \Sigma = r^2 + a^2 \cos^2 \theta,   \nonumber
\end{eqnarray}
and $M^2 \geq a^2 $. In terms of these coordinates we also have
\begin{equation}
    g_{t\varphi}^2 - g_{\varphi\varphi}g_{tt} = \Delta\sin^2 \theta \,.
\end{equation}
The Kerr metric possesses a Killing tensor field given by
\begin{eqnarray}
K_{\mu\nu} = 2\Sigma \, l_{(\mu}n_{\nu)} + r^2 g_{\mu\nu} \quad  \textrm{satisfying} \quad \nabla_{(\alpha}K_{\mu\nu)} = 0 \,, \nonumber
\end{eqnarray}
where the null vector fields $l^{\mu}$ and $n^{\mu}$ ($l^{\mu}l_{\mu}= n^{\mu}n_{\mu}=0$) satisfy the relation $l^{\mu}n_{\mu}=-1$ and read
\begin{eqnarray}
l^{\mu} &=& \frac{r^2+a^2}{\Delta} \left(\frac{\partial}{\partial t}\right)^{\mu} +
\frac{a}{\Delta} \left(\frac{\partial}{\partial \varphi}\right)^{\mu} + \left(\frac{\partial}{\partial r}\right)^{\mu}  \,, \nonumber \\
n^{\mu} &=& \frac{r^2+a^2}{2\Sigma} \left(\frac{\partial}{\partial t}\right)^{\mu} +
\frac{a}{2\Sigma} \left(\frac{\partial}{\partial \varphi}\right)^{\mu} - \frac{\Delta}{2\Sigma} \left(\frac{\partial}{\partial r}\right)^{\mu}  \,, \nonumber
\end{eqnarray}
implying the existence of a constant of motion:
\begin{eqnarray}
    \quad C = K_{\mu\nu}U^{\mu}U^{\nu} = 2\Sigma \, (l_{\mu}U^{\mu}) (n_{\mu}U^{\mu}) - r^2 = \textrm{const.} \nonumber
\end{eqnarray}
which is related to the Carter constant $Q$ \cite{Carter} as follows
\begin{eqnarray}
C &\equiv& (L-aE)^2 + Q \nonumber \\
     &=&  \frac{\left[(r^2 + a^2)E - aL \right]^2 - \Sigma^2 (U^{r})^2 - \Delta r^2}{\Delta}\,.
\end{eqnarray}
From this relation we can obtain the following expression for the radial velocity $U^{r}$:
\begin{eqnarray}
\Sigma^2 (U^{r})^2 &=& \left[(r^2\!+\!a^2)E\!-\!aL \right]^2\!-\!\Delta\left[r^2\!+\!(L\!-\!aE)^2\!+\!Q\right]
\nonumber \\
&\equiv& V^2(r)
\label{V2}
\end{eqnarray}
which is a function of the $r$ coordinate alone.
By further substituting this relation into (\ref{Ur2}) we get for the polar velocity $U^{\theta}$:
\begin{equation}
\Sigma^2(U^{\theta})^2\!=\!Q\!-\!\left[a^2(1\!-\!E^2)\!+\!\frac{L^2}{\sin^2 \theta}\right]\cos^2\theta\!\equiv\!\Theta^2(\theta),
\label{Theta2}
\end{equation}
i.e. an expression depending only on the polar angle $\theta$.
In order to give a physical interpretation to the Carter constant, we rewrite equation (\ref{Theta2}) as
\begin{eqnarray}
Q = \Sigma^2 (U^{\theta})^2 + \left[ a^2 (1-E^2) + \frac{L^2}{\sin^2 \theta} \right] \cos^2 \theta,  \nonumber
\end{eqnarray}
which gives us a measure of how much the path of the test particle departures from the equatorial plane $\theta = \pi/2$ where this
quantity vanishes.
Thus, for bounded orbits we have \cite{Bardeen,Wilkins}
\begin{equation}
\label{E<1}
    E < 1 \qquad \textrm{and} \qquad Q \geq 0,
\end{equation}
while for unbounded orbits we get
%\begin{eqnarray}
$E \geq 1\,$.   
%\nonumber
%\end{eqnarray}
%

Thus, the geodesic equations for a massive test particle with given parameters $E$, $L$, $Q$ and initial conditions $x^{\mu}_{0}$ are encoded in $U^{\mu}$ and
are given by the relations (\ref{V2})--(\ref{Theta2}) together the following expressions
\begin{eqnarray}
U^{t}\!&\!\!=\!\!&\!\frac{1}{\Delta\Sigma}\left\{\!\left[\!(\!r^2\!+\!a^2)^2\!-\!\Delta a^2 \sin^2\!\theta \right]\!E\!-\!(2Mar)\!L\right\}\!\!,   \label{Ut}
\end{eqnarray}
\begin{eqnarray}
U^{\varphi}\!&\!\!=\!\!&\!\frac{1}{\Delta\Sigma\sin^2\!\theta}\!\left[\!(2Mar\sin^2\!\theta)\!E\!+\!(\!\Delta\!-\!a^2\sin^2\!\theta)\!L\right]\!\!.  \label{Uphi}
\end{eqnarray}

It is worth noticing that for the Kerr metric, the conditions (\ref{Vmin}), (\ref{Veff=0}) and (\ref{E<1}) render the following restrictions on $r$ for circular equatorial orbits \cite{Bardeen}
\begin{equation}
\label{r_bound}
r > 2M\mp a+2\sqrt{M}\sqrt{M\pm a}.
\end{equation}
where the $\pm$ signs respectively correspond to the co-rotating and counter-rotating source (emitter or detector) with respect to the direction of the angular velocity of the Kerr black hole. Moreover, when considering equatorial circular orbits ($U^r=0=U^{\theta}$), we must impose the following condition in order to make them {\it stable}:
\begin{eqnarray}
\label{r_ge}
r &>& M\left[ 3 + Z_2 \mp \sqrt{(3-Z_1)(3+Z_1+2Z_2)}\right], \\
Z_1&\equiv&  1+\left(1-\frac{a^2}{M^2}\right)^{1/3}\left[ \left(1+\frac{a}{M}\right)^{1/3} + \left(1-\frac{a}{M}\right)^{1/3} \right] \nonumber\\
Z_2&\equiv& \sqrt{3\frac{a^2}{M^2}+Z_1^2}.   \nonumber
\end{eqnarray}

\section{Detection and emission of photons in the Kerr black hole}

Let us now consider photons with 4-momentum, parameterized by $k^{\mu} = (k^{t}, k^{r}, k^{\theta}, k^{\varphi})$, which move along null geodesics
$k^{\mu}k_{\mu} =0$ outside the event horizon of the Kerr black hole, a fact that can be expressed as
\begin{equation}
0 =  g_{tt} (k^{t})^2 + 2g_{t\varphi} (k^{t} k^{\varphi}) + g_{\varphi\varphi} (k^{\varphi})^2
    + g_{rr} (k^{r})^2 + g_{\theta\theta} (k^{\theta})^2\,.
\end{equation}
The movement of these photons is such that the following quantities are preserved:
\begin{eqnarray}
  E_{\gamma} &=& - g_{\mu\nu} \xi^{\mu} k^{\nu} = - g_{tt}k^{t} - g_{t\varphi} k^{\varphi}\,, \label{Egamma}  \\
  L_{\gamma} &=& \quad g_{\mu\nu} \psi^{\mu} k^{\nu} = g_{\varphi t}k^{t} + g_{\varphi\varphi} k^{\varphi}\,, \label{Lgamma}
\end{eqnarray}
along with a relation involving the Carter constant $Q_{\gamma}$:
\begin{eqnarray}
C_{\gamma} \equiv (L_{\gamma}-aE_{\gamma})^2 + Q_{\gamma} = K_{\mu\nu}k^{\mu}k^{\nu} =
2\Sigma \, (l_{\mu}k^{\mu}) (n_{\mu}k^{\mu}).  \nonumber
\end{eqnarray}
Therefore, the geodesic equations of photons with given parameters $E_{\gamma}$, $L_{\gamma}$, $Q_{\gamma}$ and $y^{\mu}_{o}$ are
parameterized by $k^{\mu}$ in the following way
\begin{eqnarray}
k^{t}\!=\!\frac{1}{\Delta\Sigma}\left\{\left[(r^2\!+\!a^2)^2\!-\!\Delta a^2 \sin^2\theta\right]E_{\gamma}\!-\!(2Mar)\,L_{\gamma}\right\},   \nonumber
\end{eqnarray}
\begin{eqnarray}
k^{\varphi}\!=\!\frac{1}{\Delta\Sigma \sin^2 \theta} \left[ (2Mar \sin^2 \theta)\, E_{\gamma}\!+\!(\Delta\!-\!a^2\sin^2 \theta)\, L_{\gamma} \right],  \nonumber
\end{eqnarray}
\begin{eqnarray}
\Sigma^2 (k^{r})^2\!=\!\left[(r^2\!+\!a^2)E_{\gamma}\!-\!aL_{\gamma} \right]^2\!-\!\Delta\left[(L_{\gamma}\!-\!aE_{\gamma})^2\!+\!Q_{\gamma}\right],   \nonumber
\end{eqnarray}
where the right hand side is again a function of the radial coordinate alone, and
\begin{eqnarray}
\Sigma^2 (k^{\theta})^2 &=& Q_{\gamma} - \left[ - a^2 E_{\gamma}^2 + \frac{L_{\gamma}^2}{\sin^2 \theta} \right] \cos^2 \theta\,,  \nonumber
\end{eqnarray}
where now we got a relation depending only on the polar coordinate in the right hand side.

%%%%%%%%%%%%%%%%%%%%%%%%%%%%%%%%%%%%%%%%%%%%%

\section{Red/blue shift of emitted photons}

In order to compute the red/blue shifts that emitted photons by massive particles experience while traveling along null geodesics towards an observer located far away from their source we shall mainly follow and generalize the results presented in \cite{chiapas}, where a general stationary axisymmetric metric was employed. Here we should mention that this approach analyzes the problem on the basis of the directly measured quantities: the gravitational red and blue shifts, in contrast to the tangential velocities, which are coordinate dependent observables. Moreover, this approach enables us to keep track of the effect of the underlying made assumptions, and to be aware of when they are not longer valid.

In general, the frequency of a photon measured by an observer with proper 4-velocity $U^\mu_C$ at point $P_C$ reads
\begin{equation}
\omega_C=-k_\mu U^\mu_C|_{P_C}\,,
\label{freq}
\end{equation}
where the index $C$ refers to the emission $(e)$ and/or detection $(d)$ at the corresponding space-time point $P_C$.

Thus, the frequency of light signals measured by an observer comoving with the test particle at the emission point $(e)$ is:
\begin{eqnarray}
\omega_e &=& - (k_{\mu}U^{\mu})\mid_{e}\,,   \nonumber
\end{eqnarray}
whereas the frequency detected $(d)$ by an observer located far away from the source is given by:
\begin{eqnarray}
\omega_d &=& - (k_{\mu}U^{\mu})\mid_{d}\,,  \nonumber
\end{eqnarray}
where the 4-velocities of the emitter and the detector respectively are:
\begin{equation}
\label{velocitye}
    U^{\mu}_{e} = (U^{t}, U^{r}, U^{\theta}, U^{\varphi})\mid_{e}\,,  %\nonumber
\end{equation}
\begin{equation}
\label{velocityd}
    U^{\mu}_{d} = (U^{t}, U^{r}, U^{\theta}, U^{\varphi})\mid_{d}\,.     %\nonumber
\end{equation}
In the special case in which the detector is located far enough from the source, we can consider that the observer is at infinity $(r \rightarrow \infty)$, rendering the following 4-velocity:
\begin{equation}
\label{velocityd_infty}
    U^{\mu}_{d} = (1, 0, 0, 0)\,,     %\nonumber
\end{equation}
where we have taken into account that when $r \rightarrow \infty$ the 4-velocities $U^r_d$, $U^{\theta}_{d}$ and $U^{\varphi}_{d}$ vanish, while $U^{t}_{d}$ tends to $E=1$, as it can be directly checked from (\ref{V2}), (\ref{Theta2}) and (\ref{Uphi}), and from (\ref{Ut}), respectively. Here it is worth recalling that if we indeed suppose that the orbits of the emitter are located on the equatorial plane, i.e., in the $\theta=\pi/2$ plane, meaning that the orbiting body does not move along the $\theta$-direction, then necessarily $U^{\theta}=0$ identically.

On the other hand, a photon which is emitted or detected at point $P_C$ possesses a 4-momentum
\begin{equation}
\label{kgral}
   k_C^\mu=\left(k^t,k^r,k^{\theta},k^\varphi\right)_C.
\end{equation}
Again, if the photons are considered to move along null geodesics in the equatorial plane $\theta=\pi/2$, then $k^{\theta}$ will necessarily vanish.

Thus, the frequency shift associated to the emission and detection of photons is in general given by either of the following relations:
\begin{eqnarray}
&1+z& = \frac{\omega_e}{\omega_d} \nonumber  \\
&=& \frac{(Ek^{t} - Lk^{\varphi} - g_{rr}U^{r}k^{r} - g_{\theta\theta}U^{\theta}k^{\theta})\mid_{e}}
{(Ek^{t} - Lk^{\varphi} - g_{rr}U^{r}k^{r} - g_{\theta\theta}U^{\theta}k^{\theta})\mid_{d}} \\
&=& \frac{(E_{\gamma}U^{t} - L_{\gamma}U^{\varphi} - g_{rr}U^{r}k^{r} - g_{\theta\theta}U^{\theta}k^{\theta})\mid_{e}}
{(E_{\gamma}U^{t} - L_{\gamma}U^{\varphi} - g_{rr}U^{r}k^{r} - g_{\theta\theta}U^{\theta}k^{\theta})\mid_{d}}\,,     \nonumber
\end{eqnarray}
where we have taken into account the relations (\ref{E})-(\ref{L}) for the constants $E$ and $L$ in the second line, and the relations
(\ref{Egamma})-(\ref{Lgamma}) for the constant of motion $E_{\gamma}$ and $L_{\gamma}$ in the third line, together with the frequency definition (\ref{freq}) and the expressions (\ref{velocitye}) and (\ref{velocityd}) for the 4-velocity of the emitter and the detector, respectively, as well as the relation (\ref{kgral}) for the 4-momentum of the emitted photons.

{\it This is the most general expression for the red/blue shifts that light signals emitted by massive particles experience in their path along null geodesics towards a distant observer (ideally located at spatial infinity, in particular).}

This equation for the red/blue shifts includes stable orbits of any kind for the stars and/or galactic gas: circular, elliptic, irregular, equalorial, non-equatorial, etc.

In general, the red/blue shifts represent a function $F$ of the form
\begin{eqnarray}
1 + z &=& \frac{\omega_e}{\omega_d} = F(r, \theta, E, b, B, q, s, a, M)
\label{F}
\end{eqnarray}
which is independent of the energy constant of motion of the emitted photons from the orbiting body $E_{\gamma}$; the parameters $b, B, q, s$ are defined by the following quotients:
\begin{eqnarray}
    b \equiv \frac{L_{\gamma}}{E_{\gamma}}, \quad \quad B \equiv \frac{L}{E}, \quad\quad
    q \equiv \frac{Q_{\gamma}}{E^2_{\gamma}}, \quad\quad s \equiv \frac{Q}{E^2}.    \nonumber
\end{eqnarray}

Here we should stress that there are two different frequency shifts which correspond to the maximum and minimum values of $\omega_e$ related
to the light propagation along the same and the opposite direction with respect to the motion of the emitter of photons orbiting a black hole, i.e., the frequency shifts corresponding to a receding (red shift) and to an approaching (blue shift) photon source, respectively. These maximum and minimum values of the frequency shift are reached for bodies whose position vector ${\bf r}$, with respect to the black hole center, is orthogonal to the detector's line of sight, i.e., along the plane where $k^r$ vanishes for an observer located far away from the source of light signals.

%%%%%%%%%%%%%%%%%%%%%%%%%%%%%%%%%%%%%%%%%%%%%%%%%%%%%%%%%%%%%%%%%%%%%%%%%%%%%%%%%

\section{The red/blue shift of photons in circular and equatorial orbits around the Kerr black hole}

We shall further restrict ourselves to the study of circular and equatorial orbits ($U^{r}=U^{\theta}=0$) to give an example of how the above developed formalism can be applied to the determination of the mass and rotation parameters of the Kerr black hole from the measured red/blue shift of photons detected far away from the source.

In this case the expression for the red/blue shift of light signal becomes
\begin{equation}
\label{zcircorbits}
1+z =
\frac{\omega_e}{\omega_d} = \frac{\left.\left(E_\gamma U^t-L_\gamma
U^\varphi\right)\right|_e}{\left.\left(E_\gamma U^t - L_\gamma U^\varphi\right)\right|_d} =
\frac{U^t_e - b_e \,U^\varphi_e}{U^t_d - b_d \,U^\varphi_d}\,\,\,\,,
\end{equation}
where we introduced the apparent impact parameter $b\equiv\frac{L_\gamma}{E_\gamma}$, where $E_\gamma$ and $L_\gamma$ are defined by (\ref{Egamma}) and (\ref{Lgamma}), respectively. Since the constants of motion $E_\gamma$ and $L_\gamma$ are preserved along the null geodesics followed by the photons from emission till detection, therefore $b_e=b_d$, i.e. this quantity is also constant along the whole photons path. Here we should note that this relation is quite important since it links the observed radius of star/gas orbits with the radius of the observers orbit, i.e. with the distance to the black hole source (see below).

We shall further consider the kinematic red/blue shifts of photons either side of the central value $b=0$, this renders two values for $b$ different in magnitude which are generated by the rotational character of the gravitational field, see (\ref{b}) below. In order to accomplish this, we need to compute the gravitational red shift corresponding to a photon emitted by a static particle located at $b=0$:
\begin{equation}
\label{zatbnull}
1+z_c = \frac{U^t_e}{U^t_d}
%=\frac{\left.\left(\frac{E-LW}{e^{2\Phi}}\right)\right|_e}{\left.\left(\frac{E-LW}{e^{2\Phi}}\right)\right|_d}
\,\,,
\end{equation}
and to subtract this quantity from (\ref{zcircorbits}) in order to define the kinematical red shift
\begin{equation}
\label{zkin}
z_{\textrm{kin}}\equiv z-z_c = \frac{U^t_e U^\varphi_d b_d - U^t_d U^\varphi_e b_e}
{U^t_d\left(U^t_d - b_d\,U^\varphi_d\right)}\,.
\end{equation}
Actually, this analysis can be performed in the same way without subtracting the magnitud $z_c$ from $z$. Of course, this change will modify the obtained values for $z_1$ and $z_2$ (see below). However, some astronomers report their data in terms of the kinematical redshifts, i.e. with the redshift of the galaxy subtracted from $z$.

We further need to take into account the light bending due to the gravitational field generated by
the rotating black hole, in other words, we need to construct a mapping between the apparent impact parameter
$b$ and the location of the emitter $r$ given by its vector position ${\bf r}$ with respect to the
center of the source, i.e., the mapping $b(r)$. Following \cite{nucamendi,lake}, we shall choose
the maximum value of $z$ at a fixed distance from the observed center of the source (at a fixed
$b$). From (\ref{zkin}) it follows that if the prefactor that multiplies $b$ is a
monotonically decreasing function with increasing $r$, then the maximum observed value of $z_{\textrm{kin}}$ corresponds to
the minimum value of $r$ along the null geodesic of the photons. This minimum value of $r$ corresponds
to the position of the orbiting object either side of the center of the source, where the photon is emitted with a $k^r=0$ component.

Thus, from the expression (\ref{zkin}), it follows that the apparent impact parameter $b$ must also be maximized; this quantity can be calculated from the geodesic equation of the photons (or, equivalently, from the $k^\mu k_\mu =0$ relation taking into account that $k^r=0$ and $k^\theta=0$) and is given by
\begin{equation}
\label{b}
b_\pm = - \frac{g_{t\varphi}\pm\sqrt{g_{t\varphi}^2-g_{tt}g_{\varphi\varphi}}}{g_{tt}} \,,
\end{equation}
where we got two values, $b_-$ and $b_+$ (either evaluated at the emitter
or detector position, since this quantity is preserved along the null geodesic trajectories of the photons, i.e. $b_e=b_d$) that respectively give rise to two different shifts, $z_1$ and $z_2$, of the emitted photons corresponding to a receding and to an approaching object with respect to a far away positioned observer:
\begin{equation}
\label{z+}
z_1 = \frac{U^t_e U^\varphi_d\, b_{d_-} - U^t_d U^\varphi_e\, b_{e_-}}{U^t_d\left(U^t_d - U^\varphi_d\, b_{d_-}\right)}\,,
\end{equation}
\begin{equation}
\label{z-}
z_2 = \frac{U^t_e U^\varphi_d\, b_{d_+} - U^t_d U^\varphi_e\, b_{e_+}}{U^t_d\left(U^t_d - U^\varphi_d\, b_{d_+}\right)}\,.
\end{equation}

In general, $|z_1| \neq |z_2|$ as can be easily seen from (\ref{b})-(\ref{z-}) because of two reasons: the light bending experienced by the emitted photons either side of the geometrical center of the source, and the differential rotation experienced by the detector codified by $U^{\varphi}_{d}$ and $U^{t}_{d}$. In fact, the second term in the denominator of (\ref{z+}) and (\ref{z-}) encodes
the contribution of the movement of the detector's inertial frame. If this quantity is negligible in comparison to the contribution coming from the
$U_d^t$ component ($U_d^\varphi<<U_d^t$), then the detector can be considered static at spatial infinity. Let us define
\begin{equation}
\label{OmegaD}
\frac{U_d^\varphi}{U_d^t}=\frac{d\varphi}{dt}\equiv \Omega_d,
\end{equation}
as the angular velocity of a detector located far away from the photons source. Thus, when this quantity is small, $\Omega_d<<1$, the detector can be treated as static, neglecting its relative movement. In terms of $\Omega_d$, the $z_1$ and $z_2$ read:
\begin{equation}
\label{z++}
z_1 = \frac{U^t_e\, \Omega_d\, b_{d_-} - U^\varphi_e\, b_{e_-}}{U^t_d\left(1 - \Omega_d\, b_{d_-}\right)}\,,
\end{equation}
\begin{equation}
\label{z--}
z_2 = \frac{U^t_e\, \Omega_d\, b_{d_+} - U^\varphi_e\, b_{e_+}}{U^t_d\left(1 - \Omega_d\, b_{d_+}\right)}\,.
\end{equation}

It is easy to see that when $\Omega_d<<1$, the $z_1$ and $z_2$ are still different in magnitude, since these quantities are respectively proportional to $b_{e_-}$ and $b_{e_+}$.

When we drop indeed the rotational metric component $g_{t\varphi}$ in (\ref{b}) we recover the spherically symmetric light bendings given by $b_{\pm}=\pm\sqrt{-g_{\varphi\varphi}/g_{tt}}$, which differ each other in sign but possess the same magnitude.

Thus, the gravitational rotation bends the light in a different way for approaching and receding photon sources.

In order to get a close expression for the gravitational red/blue shifts experienced by the emitted photons we shall express the required quantities in terms of the Kerr black hole metric. Thus, the $U^{\varphi}$ and $U^{t}$ components of the 4-velocity for circular equatorial orbits read
\begin{eqnarray}
U^{\varphi}(r, \pi/2) &=& \frac{(2Ma)E + (r-2M)L}{r (r^2 + a^2 - 2Mr)}\,, %\nonumber
\label{UphiKerr}
\end{eqnarray}
\begin{eqnarray}
U^{t}(r, \pi/2) &=& \frac{(r^{3} + a^{2}r + 2Ma^{2})E - (2Ma)L}{r (r^2 + a^2 - 2Mr)}\,,  %\nonumber
\label{UtKerr}
\end{eqnarray}
whereas the constants of motion $E$ and $L$ are
\begin{equation}
E = \frac{r^{3/2} - 2Mr^{1/2} \pm aM^{1/2}}{r^{3/4} \, (r^{3/2} - 3Mr^{1/2} \pm 2aM^{1/2})^{1/2}}\,,   %\nonumber
\label{EKerr}
\end{equation}
\begin{equation}
L = (\pm) \frac{M^{1/2} \, (r^{2} \mp 2aM^{1/2}\, r^{1/2} + a^{2})}{r^{3/4} \, (r^{3/2} - 3Mr^{1/2} \pm 2aM^{1/2})^{1/2}}\,.   %\nonumber
\label{LKerr}
\end{equation}
where the $\pm$ signs again correspond to the co-rotating and counter-rotating objects (either the emitter or the detector) with respect to the direction of the angular velocity of the Kerr black hole \cite{Bardeen}. By substituting (\ref{EKerr}) and (\ref{LKerr}) into the expressions for (\ref{UphiKerr}) and (\ref{UtKerr}) we finally obtain for the latter quantities
\begin{eqnarray}
U^{\varphi}(r, \pi/2) &=& \frac{\pm M^{1/2}}{r^{3/4} \sqrt{r^{3/2} - 3Mr^{1/2} \pm 2aM^{1/2}}}\,, %\nonumber
\label{UphiKerr2}
\end{eqnarray}
\begin{eqnarray}
U^{t}(r, \pi/2) &=& \frac{\left(r^{3/2} \pm a M^{1/2}\right)}{r^{3/4} \sqrt{r^{3/2} - 3Mr^{1/2} \pm 2aM^{1/2}}}\,.  %\nonumber
\label{UtKerr2}
\end{eqnarray}
With these quantities at hand it is straightforward to compute the angular velocity of a source orbiting around the Kerr black hole
\begin{equation}
\label{Omega}
\Omega_{\pm}=\frac{\pm M^{1/2}}{\left(r^{3/2} \pm a M^{1/2}\right)}\,
\end{equation}
in a circular and equatorial orbit. It is worth mentioning that this angular velocity corresponds to either the emitter or the detector of photons, in which case the subscripts $_e$ and $_d$ must be respectively used (see definition (\ref{OmegaD}) for the detector case). The $\pm$ signs correspond to co-rotating and counter-rotating objects with the angular velocity of the Kerr black hole.

On the other side, for the Kerr black hole metric, equation (\ref{b}) renders the following expression for the mapping $b(r)$, responsible for the gravitational light bending, for circular and equatorial orbits:
\begin{equation}
\label{bKerr}
b_{\pm} =\frac{-2aM \pm r \sqrt{r^2 + a^2 - 2Mr}}{r - 2M} \,.
%\nonumber
\end{equation}
where we have taken into account its maximum character, i.e. the fact that the photons are {\it emitted} at the point where $k^r=0$.
It turns out that the quantities $z_1$ y $z_2$ correspond to the red and blue shifts, $z_{red}$ and $z_{blue}$, respectively, according to the plots of $z_1$ and $z_2$ in terms of the radial coordinate $r$ and the Kerr black hole parameters $M$ and $a$.

Therefore, for the Kerr black hole case we can write the red and blue shifts (\ref{z++}) and (\ref{z--}), respectively, as
\begin{eqnarray}
\label{ZrKerr}
z_{red}\!=\!\frac{r_d^{\frac{3}{4}}\sqrt{r_d^{\frac{3}{2}}\!-\!3Mr_d^{\frac{1}{2}}\!\pm\!2aM^{\frac{1}{2}}}\,\Omega_{d_{\pm}}\!
\left(\Omega_{d_{\pm}}\,b_{d_-}\!-\Omega_{e_{\pm}}\, b_{e_-}\right)}
{r_e^{\frac{3}{4}}\sqrt{r_e^{\frac{3}{2}}\!-\!3Mr_e^{\frac{1}{2}}\pm2aM^{\frac{1}{2}}}\,\,
\Omega_{e_{\pm}}\left(1-\Omega_{d_{\pm}}\, b_{d_-}\right)}\,,
\nonumber
\end{eqnarray}
\begin{eqnarray}
z_{blue}\!=\!\frac{r_d^{\frac{3}{4}}\sqrt{r_d^{\frac{3}{2}}\!-\!3Mr_d^{\frac{1}{2}}\!\pm\!2aM^{\frac{1}{2}}}\,\Omega_{d_{\pm}}\!
\left(\Omega_{d_{\pm}}\,b_{d_+}\!\!-\Omega_{e_{\pm}}\, b_{e_+}\right)}
{r_e^{\frac{3}{4}}\sqrt{r_e^{\frac{3}{2}}\!-\!3Mr_e^{\frac{1}{2}}\pm2aM^{\frac{1}{2}}}\,\,
\Omega_{e_{\pm}}\left(1-\Omega_{d_{\pm}}\, b_{d_+}\right)}\,,
\nonumber
\label{ZbKerr}
\end{eqnarray}
where now $r_e$ and $r_d$ stand for the radius of the emitter's and detector's orbits, respectively, and the $ _{\pm}$ subscripts correspond, as before, to the co-rotating and counter-rotating source with respect to the direction of the angular velocity of the Kerr black hole.

These expressions can be written as well in terms of the Kerr black hole parameters, $M$ and $a$, and the detector radius, $r_d$, as follows
\begin{eqnarray}
\label{ZrK}
&&z_{red}\!=\!\pm M^{\frac{1}{2}}\, \frac{r_d^{\frac{3}{4}}\sqrt{r_d^{\frac{3}{2}}\!-\!3Mr_d^{\frac{1}{2}}\pm2aM^{\frac{1}{2}}}\,
\left(r_d^{\frac{3}{2}}\!-\!r_e^{\frac{3}{2}}\right)}
{r_e^{\frac{3}{4}}\sqrt{r_e^{\frac{3}{2}}\!-\!3Mr_e^{\frac{1}{2}}\pm2aM^{\frac{1}{2}}}\,
\left(r_d^{\frac{3}{2}}\!\pm\!aM^{\frac{1}{2}}\right)}\times \nonumber \\
&&
\frac{\left(2aM + r_e\sqrt{r_e^2 - 2Mr_e + a^2}\right)}
{\left[r_d^{\frac{3}{2}}\!\left(r_e\!\!-\!\!2M\right)\!\pm\!aM^{\frac{1}{2}}r_e\!\pm\!M^{\frac{1}{2}}r_e\sqrt{r_e^2\!\!-\!\!2Mr_e\!\!+\!a^2}\right]},
\end{eqnarray}
\begin{eqnarray}
\label{ZbK}
&& z_{blue}\!=\!\pm M^{\frac{1}{2}}\frac{r_d^{\frac{3}{4}}\sqrt{r_d^{\frac{3}{2}}\!-\!3Mr_d^{\frac{1}{2}}\pm2aM^{\frac{1}{2}}}\,
\left(r_d^{\frac{3}{2}}\!-\!r_e^{\frac{3}{2}}\right)}
{r_e^{\frac{3}{4}}\sqrt{r_e^{\frac{3}{2}}\!-\!3Mr_e^{\frac{1}{2}}\pm2aM^{\frac{1}{2}}}\,
\left(r_d^{\frac{3}{2}}\!\pm\!aM^{\frac{1}{2}}\right)}\times \nonumber \\
&&
\frac{\left(2aM - r_e\sqrt{r_e^2 - 2Mr_e + a^2}\right)}
{\left[r_d^{\frac{3} {2}}\!\left(r_e\!\!-\!\!2M\right)\!\pm\!aM^{\frac{1}{2}}r_e\!\mp\!M^{\frac{1}{2}}r_e\sqrt{r_e^2\!\!-\!\!2Mr_e\!\!+\!a^2}\right]},
\end{eqnarray}
where the radii of the orbits of stars, denoted by $r_e$, are given data obtained from observations, and we have made use of the relation $b_e=b_d$.

Here we should point out that in the special case of circular and equatorial orbits around the Kerr black hole, the red and blue shifts now constitute simplified functions $F(r_d, a, M)$ in comparison to (\ref{F}).

Remarkably, from the fact that the constants of motion $E_{\gamma}$ and $L_{\gamma}$, and hence the apparent impact parameter $b$, are preserved along the whole trajectory followed by photons, the latter quantity is the same when evaluated either at the emitter or detector position, rendering the following relation $b_e=b_d$. Therefore, this equation links the emitter and detector radii: 
\begin{eqnarray}
\label{beesbd}
&& r_d = \sqrt{\frac{b_e-a}{3}}\times \\
&& \left[\!\left(\!\sqrt{27M^2(b_e\!-\!a)\!+\!(b_e\!+\!a)^3}\!-\!\sqrt{27M^2(b_e\!-\!a)}\right)^{\frac{1}{3}}\!+\! \right.   \nonumber\\
&& \left.\!(b_e\!+\!a)\!\left(\!\sqrt{27M^2(b_e\!-\!a)\!+\!(b_e\!+\!a)^3}\!-\!\!\sqrt{27M^2(b_e\!-\!a)}\right)\!^{\!-\!\frac{1}{3}}\!\right]\!\nonumber
\end{eqnarray}
rendering an expression for calculating $r_d$ once the emitter radius has been measured and the mass and rotation parameters have been estimated.

{\it Thus, for a given set of constant data $r_e$ which characterizes the radius of circular paths of orbiting emitters (stars and galactic gas/dust) around the Kerr black hole, together with the measured red and blue shifts, $z_{red}$ and $z_{blue}$, experienced by the emitted photons at the points where $b_{\pm}$ is maximized, we can determine the mass and rotation parameters, $M$ and $a$, of the Kerr black hole as well as the distance from the detector to the source codified by $r_d$, through the exact relations (\ref{ZrK}) and (\ref{ZbK}).}

In the particular case when the detector is located far away from the source and the following condition is fulfilled $r_d>>M\ge a$, the red and blue shifts respectively become
\begin{eqnarray}
\label{ZrKstatic}
z_{red} = \frac{\pm M^{\frac{1}{2}}\left(2aM + r_e\sqrt{r_e^2 - 2Mr_e + a^2\,}\right)}
{r_e^{\frac{3}{4}}\,\big(r_e-2M\big)\sqrt{r_e^{\frac{3}{2}}\!-\!3Mr_e^{\frac{1}{2}}\pm2aM^{\frac{1}{2}}\,}},
\end{eqnarray}
\begin{eqnarray}
z_{blue} = \frac{\pm M^{\frac{1}{2}}\left(2aM - r_e\sqrt{r_e^2 - 2Mr_e + a^2\,}\right)}
{r_e^{\frac{3}{4}}\,\big(r_e-2M\big)\sqrt{r_e^{\frac{3}{2}}\!-\!3Mr_e^{\frac{1}{2}}\pm2aM^{\frac{1}{2}}\,}}.
\label{ZbKstatic}
\end{eqnarray}
In this special case the mass and rotation parameters of the Kerr black hole can be obtained from the measured red and blue shifts of the photons emitted by the stars through equations (\ref{ZrKstatic}) and (\ref{ZbKstatic}). In principle, one could be tempted to algebraically express the mass and rotation parameter from the latter equations. However, even though it is easy to extract a closed expression for the rotation parameter
\begin{eqnarray}
\label{a2}
a^2= \frac{r_e^3\left(r_e-2M\right)\left(z_{red}+z_{blue}\right)^2}
{4M^2\,\,\left(z_{red}-z_{blue}\right)^2 - r_e^2\left(z_{red}+z_{blue}\right)^2},
\nonumber
\end{eqnarray}
the equation to solve for the mass is of eighth order and cannot be solved exactly:
\begin{eqnarray}
\left[16\, r_e M^3-\left(4\beta M^2-\alpha r_e^2\right)\left(r_e-2M\right)\left(r_e-3M\right) \right]^2 = \nonumber \\
4\alpha r_e^2 M\left(r_e-2M\right)^3\left(4\beta M^2-\alpha r_e^2\right),
\label{M8}
\nonumber
\end{eqnarray}
where we have introduced $\alpha \equiv \left(z_{red}+z_{blue}\right)^2$ and $\beta \equiv \left(z_{red}-z_{blue}\right)^2$.  Thus, one must turn to a Bayesian fitting in order to compute this quantities from observational data.

%%%%%%%%%%%%%%%%%%%%%%%%%%%%%%%%%%%%%%%%%%%%%%%%%%%%%%%%%%%%%%%%%%%%%%%%%%%%%%%%%

\subsection{More general cases}

The apparent impact parameter $b$ for the Kerr black hole family can also be calculated in the case in which the considered orbits depart from the equatorial plane and, hence, $\theta\ne\pi/2$; again, this quantity is computed from the $k^\mu k_\mu =0$ relation just taking into account its maximum character, i.e. that $k^r=0$, and reads
\begin{equation}
\label{bKerrtheta}
b_\pm =\frac{-2aMr \pm \sqrt{\Delta\left[r^4 - q \left(r^2 - 2Mr\right)\right]}}{r^2 - 2Mr} \,,
%\nonumber
\end{equation}
which renders the expression (\ref{bKerr}) when the Carter constant $Q_\gamma$ (or $q$) vanishes.

It is worth mentioning that recently published observational data (see \cite{Morris} for instance) reveal that the orbits of stars seem to be ellipses which do not lie in the equatorial plane and therefore the physically relevant problem involves calculations of the red/blue shifts with $\theta\ne\pi/2$.

%%%%%%%%%%%%%%%%%%%%%%%%%%%%%%%%%%%%%%%%%%%%%%%%%%%%%%%%%%%%%%%%%%%%%%%%%%%%%%%%%

\section{Discussion and final remarks}

In this paper we have shown that the relativistic stationary axisymmetric formalism previously constructed for the galactic rotation curve's problem presented in \cite{chiapas} can also be applied to the study of black hole rotation curves, leaving the galactic framework considered in that work.

As an application of this formalism, we can determine the Kerr black hole parameters $M$ and $a$ (and hence its angular momentum) as well as the radius of the detector's orbit $r_d$ in terms of the red and blue shifts of photons that travel along null geodesics and are emitted by massive bodies orbiting around the black hole following arbitrary paths (\textit{black hole rotation curves}) like the recently reported closed orbits around the center of our galaxy \cite{Morris}, for instance.

As an explicit example, we computed the red and blue shifts experienced by photons emitted by massive objects orbiting the Kerr black hole in equatorial and circular orbits and following null geodesics towards a distant observer.

Moreover, the aforementioned expressions for the red/blue shifts allow one to statistically estimate the Kerr black hole parameters $M$ and $a$ as well as the radius of the detector's orbit $r_d$ by means of a Bayesian fitting, i.e., by giving random data for the red/blue shifts and their respective errors. This analysis will allow us to know the level of precision which is required to measure the red/blue shifts in a real experiment. We really do hope to perform this analysis in the near future.

We should also mention that this formalism can be applied as well to a wider range of astrophysical phenomena like accretion discs of rotating black holes, binary systems and active galactic nuclei where the magnitude of the effects are less restrictive in comparison to the rotation curves around the center of our galaxy.

As a further research line, we can generalize the present approach to the case of more general orbits: non-equatorial circular paths, elliptic equatorial orbits, elliptic non-equatorial ones, non-elliptic trajectories, etc. This situation is significantly important to characterize in a relativistic way the putative black hole hosted at the center of our galaxy on the basis of measured orbital data; more generally, this research will shed more light on the hypothesis of the existence of supermassive black holes in the center of several galaxies.

%%%%%%%%%%%%%%%%%%%%%%%%%%%%%%%%%%%%%%%%%%%%%%%%%%%%%%%%%%%%%%%%%%%%%%%%%%%%%%%%%

\section*{Acknowledgements}
Authors acknowledge illuminating correspondence and discussions with D. Sudarsky, R. Cartas--Fuentevilla, S. Gillessen and A.M. Ghez. AHA is grateful to the staff of the Physics Department of UAM-I for warm hospitality.
This research was supported by grants CIC-UMSNH, VIEP-BUAP-HEAA-EXC15-I and
\textit{Programa de Apoyo a Proyectos de Investigaci\'on e Innovaci\'on Tecnol\'ogica} (PAPIIT) UNAM, IN103413-3, \textit{Teor\'ias de Kaluza-Klein, inflaci\'on y perturbaciones gravitacionales}. Both authors thank SNI and PROMEP-SEP for support.

%%%%%%%%%%%%%%%%%%%%%%%%%%%%%%%%%%%%%%%%%%%%%%%%%%%%%%%%%%%%%%%%%%%%%%%%%%%%%%%%%

\end{document}